\documentclass[10pt,nofootinbib,superscriptaddress]{article}
\usepackage{color} 

\usepackage{amsbsy}
\usepackage{amscd}
\usepackage{amsopn}
\usepackage{amsxtra}
\usepackage{graphicx}
\usepackage{epsfig}
\usepackage{enumitem} 
\usepackage{url}
\usepackage{epstopdf}
\usepackage{authblk}
\usepackage{hhline}
\usepackage{amsmath,amstext,amsgen,amsfonts}
\def\t{\tau}
\def\O{\Omega}

\def\dd{\mathrm{d}}
\newcommand{\quotes}[1]{``#1''}
\newcommand*{\pd}[3][]{\ensuremath{\frac{\partial^{#1} #2}{\partial #3}}}

\graphicspath{ {Images/} }

\begin{document}
	
	\title{{Spreading of Memes on Multiplex Networks}}
	\author[1]{Joseph D.~O'Brien}
	\author[2]{Ioannis K.~Dassios}
	\author[1]{James P.~Gleeson}
	\affil[1]{MACSI, Department of Mathematics and Statistics, University of Limerick, Ireland}
	\affil[2]{AMPSAS, University College Dublin, Ireland}
	\date{}

	\maketitle
\begin{abstract}
A model for the spreading of online information or ``memes'' on multiplex networks is introduced and analyzed using branching-process methods. The model generalizes that of [Gleeson et al., Phys. Rev. X., 2016] in two ways. First, even for a monoplex (single-layer) network, the model is defined for any specific network defined by its adjacency matrix, instead of being restricted to an ensemble of random networks. Second, a multiplex version of the model is introduced to capture the behavior of users who post information from one social media platform to another. In both cases the branching process analysis demonstrates that the dynamical system is, in the limit of low innovation, poised near a critical point, which is known to lead to heavy-tailed distributions of meme popularity similar to those observed in empirical data.
\end{abstract}

\section{Introduction}

The advent of social media and the resulting ability to instantaneously communicate ideas and messages to connections worldwide is one of the great consequences arising from the telecommunications revolution over the last century. Individuals do not, however, communicate only upon a single platform; instead there exists a plethora of options available to users, many of whom are active on a number of such media. While each platform offers some unique selling point to attract users, e.g.,  keeping up to date with friends through messaging and statuses (\textit{Facebook}), photo sharing (\textit{Instagram}), seeing information from friends, celebrities and numerous other outlets (\textit{Twitter}) or keeping track of the career paths of friends and past colleagues (\textit{Linkedin}), the platforms are all based upon the fundamental mechanisms of connecting with other users and transmitting information to them as a result of this link.

The dynamics of information flow in  online social media is an active research area (see, for example, references \cite{bakshy2011everyone,lerman2012social,weng2012competition,cheng2014can,gleeson2014simple,weng2013virality,gleeson2014competition,notarmuzi2018analytical}) but analytically solvable models are relatively rare. An analytically tractable
 model for the spreading of `memes' (the term is used here in the general sense of a piece of digital information) on a single platform for a  directed network was introduced in \cite{gleeson2016effects}. This model suggested that as a result of the large amount of information received by users on such platforms competing for their limited attention, the system is poised near criticality.
 The implications of the near-criticality of such systems include fat-tailed distributions of meme popularity, similar to those observed in empirical studies \cite{weng2012competition}. The model also incorporated memory effects by allowing users to look backwards a random amount of time  in their stream or feed to `retweet' a meme, which gives a non-Markovian model for this aspect of human behavior \cite{jo2014analytically, iribarren2011branching}. Interestingly, while this was a null model in the sense that it was entirely derived from first principles with no assumptions being made regarding the parameters determining the spreading process, it could accurately describe observed cascades of `tweet' popularities on the directed network Twitter.
 
 The model of  \cite{gleeson2016effects} was, however, overly simplistic in several respects. Firstly, the model was restricted to ``degree-class'' networks, meaning that it applies to an ensemble of random networks in which all users with the same numbers of connections (i.e., same in-degree and out-degree) are considered to behave in the same way, with random connections between users. Because of this assumption, important parameters of the model such as the innovation probability and the distribution of memory times
were assumed to be the same for all users. A natural question (we address here) is whether the results of  \cite{gleeson2016effects} remain valid if these restrictive assumptions are relaxed. Secondly, only a monoplex (single-platform) network was considered in  \cite{gleeson2016effects} so that the effects of users having a presence on multiple platforms was ignored. In reality, the connections of users may vary on a platform-by-platform basis depending upon their use of each site, and it is important to consider the possibility of users sharing content between platforms. To study such spreading processes we require the theory of multiplex networks \cite{kivela2014multilayer}, which has previously been used to model processes as varied as the spreading of infectious diseases \cite{cozzo2013contact}, opinion dynamics \cite{amato2017opinion},  neural dynamics \cite{nicosia2017collective, vaiana2018multilayer} and dependencies among financial time series \cite{musmeci2017multiplex}.

In this article we will extend the model described in \cite{gleeson2016effects} in two ways. Firstly we shall consider the same dynamics but on a specific network rather than a random network, meaning that the full detail of the specific adjacency matrix can be incorporated, and that user-specific parameters can be defined. We will show that in the small innovation, large time limit that such a system approaches criticality. Second, we will extend the model to a multiplex framework, where each layer represents a different social media platform, we will then allow users to share content between their accounts on these different platforms and investigate the effects upon the criticality of the system. \cite{adami2002critical}
The remainder of the paper is structured as follows. Section \ref{model} describes the extended model for a multiplex specific network (including the monoplex specific network as a special case). Section \ref{criticality}  discusses the effect that the parameters in this model have on the criticality of the system before confirming the results using numerical simulations, and finally, in Section \ref{discussion}, we  discuss our results.

\section{Model}\label{model}

The model of  \cite{gleeson2016effects} considers a directed network where each node represents an individual user while an edge from user $j$ to user $i$ implies that $i$ follows $j$. The key component of this model is that each user has a stream which is a record of all `tweets' they have ever received. Memes spread in this model by users `tweeting', which may occur in two  ways. Firstly the user may innovate and create a new unique meme with probability $\mu$, which enters their own stream and also is viewed by their followers, each of whom find it interesting, and thus accept it into their stream, with probability $\lambda$. Secondly, the user may decide to `retweet' content they have previously accepted into their own feed with probability $1 - \mu$, the meme they decide to retweet is determined by their `memory time' distribution $\Phi(t_m)$ such that the random amount of time which they look back is drawn from $\Phi(t_m)$ and the meme that was present in their stream at time $t - t_m$ is chosen, where $t$ is the current time. Also, the rate at which a user tweets is determined by the activity rate $\beta$, which is common to all users with the same in- and out-degrees. We refer the reader to \cite{gleeson2016effects} for a more detailed explanation of the model.

We wish to extend the above model by considering a multiplex network consisting of $N$ nodes and $M$ layers, such that each layer represents a different social media platform while each node represents a unique account of a user. Throughout this article we assume that the content sharing mechanism of each platform is identical. Our focus is on the specific network case whereby we know the exact network topology and thus have the adjacency matrix $A = \oplus_{\alpha}A_{\alpha}$, where $A_{\alpha}$ represents the adjacency matrix of layer $\alpha$. We note in passing that the case where $M = 1$ will exactly describe the specific network extension of the degree-class network model used in \cite{gleeson2016effects}. The aim of our model is to allow users to share content which they have seen from their account on one social media platform to another of their accounts on an alternative platform, this requires us to an interlayer coupling matrix $C$ with elements equal to 1 if the nodes $i$ and $j$ represent accounts of the same user on two different platforms \cite{kivela2014multilayer}, and zero otherwise. Such notation allows our model to incorporate individuals owning accounts on multiple different platforms and also allows for each layer to have a different number of accounts: both these characteristics are desirable features for such a model and commensurate with empirical findings \cite{chaffey2016global}.

\begin{figure}[t]
	\centering
	\includegraphics[width=0.7\linewidth]{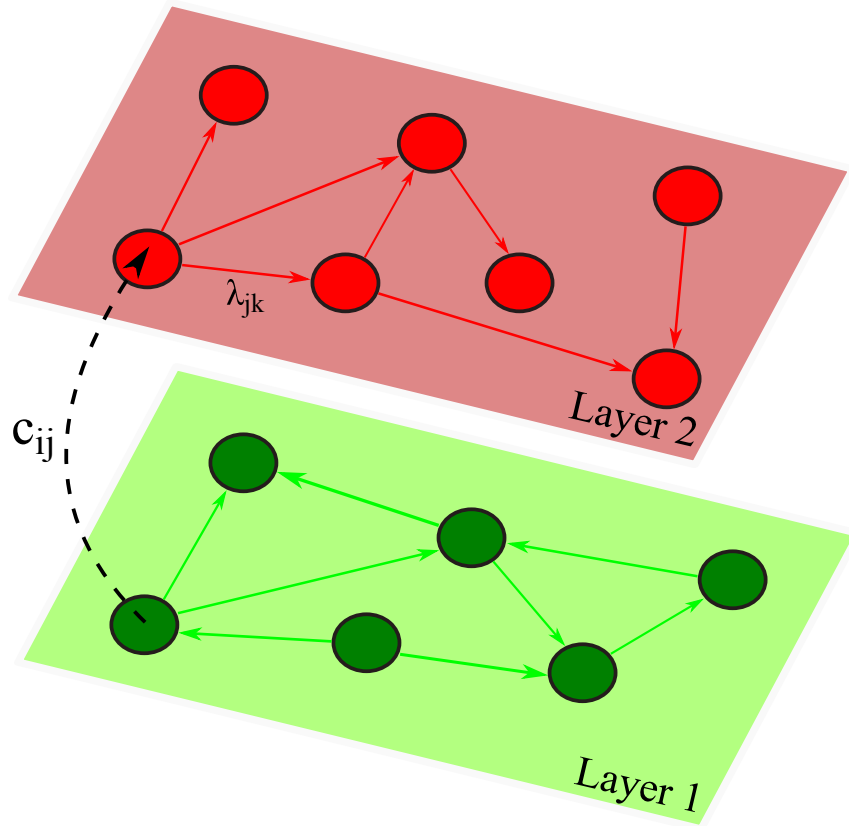}
	\caption{Schematic showing how memes may spread in our multiplex model, specifically we show an account $i$ on layer 1 sharing a meme to another account $j$ on layer 2 owned by the same user with probability $c_{ij}$, who then immediately shares it with their own followers each of whom may or may not find it interesting. The arrows indicate the direction of information flow (i.e., arrows point from followed node to following node).}
	\label{fig:multiplex}
\end{figure}

In our model these matrices $A$ and $C$ will not be used directly themselves but rather are proxies for two alternative matrices. The first of these is $\Lambda$ which has elements $\lambda_{ij}$ representing the probability that the account $j$ finds content from account $i$ interesting and thus accepts it into their stream for possible future retweeting. This matrix is directly related to $A$ as $\lambda_{ij} \ne 0$ if and only if user $j$ is connected to user $i$ on the same layer. Similarly we have a matrix $\widetilde{C}$ with elements $c_{ij}$ describing the probability that content is shared from an account $i$ to an account $j$ on a different layer, which in our model may only occur if the accounts are owned by the same user and as such $\widetilde{C}$ is a direct relation of $C$. The dynamics of the model is such that a user may see a meme they find interesting while using their accounts on one platform e.g.~Twitter, and decide to share it through their account on another platform e.g.~Instagram. The main characteristics governing the spreading in our model are highlighted in Fig.~\ref{fig:multiplex}, where node $i$ shares a meme from layer 1 to the node $j$ in layer 2 (both of which are accounts owned by the same individual), this account then immediately proceeds to share it to their followers one of whom, account $k$, will accept it into their stream with probability $\lambda_{jk}$.

\begin{figure}[b!]
	\centering
	\includegraphics[width=0.7\linewidth]{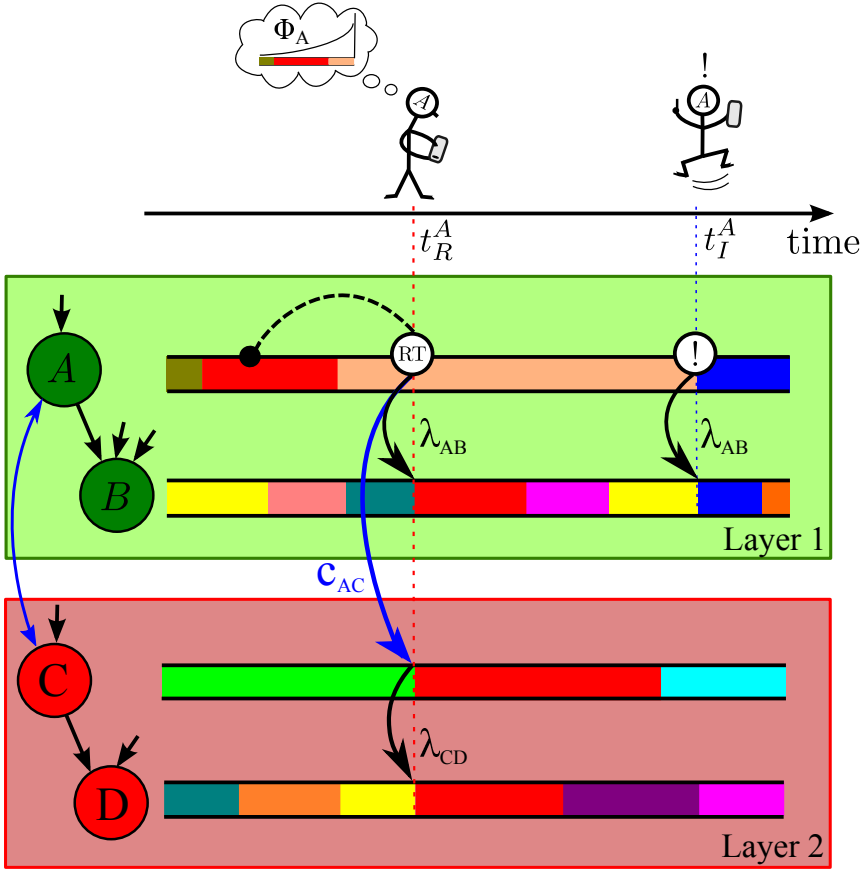}
	\caption{Schematic of the model. We consider the possible ways in which account A that is on layer 1 may spread memes in our model, firstly the different colors represent the memes present in each accounts stream, and at time $t_R^A$ account A decides to retweet a previously seen meme. She chooses the red meme by looking back in her stream a time determined by her memory-time distribution $\Phi_A$ and it is shared to each of her followers on layer 1 one of whom, user B, finds it interesting with probability $\lambda_{AB}$ and thus accepts it into their stream. Account C on a different platform is owned by the same user (highlighted by the blue arrow joining them) and as a result there is a probability $c_{AC}$ that A shares the meme across platform, if this occurs account C immediately shares it with all their followers on layer 2 one of whom, user D, accepts this meme into their stream with probability $\lambda_{CD}$. At time $t_I^A$ user A decides the innovate and create a new unique meme which is immediately seen by all their followers only on layer 1, unless at some point in the future they decide to share it across layers as before.}
	\label{fig:memory}
\end{figure}

This framework also allows us (in contrast to \cite{gleeson2016effects})  to define user-specific parameters, namely for a given node $i$ we have their innovation probability $\mu_i$, i.e., the probability that the user creates a new meme from their account and immediately shares it with their followers (this meme also appears in their own stream as they find their own content automatically interesting). The second quantity that is specific to each account is their activity rate $\beta_i \ge 0$, which determines how often they become active either to innovate or to retweet a previously seen meme from their stream. Finally, if the user does decide to retweet on their account at time $t$ then they will look backwards in their stream to a time $t - t_m$, where $t_m$ is a random variable determined by their \quotes{memory-time} distribution $\Phi_i(t_m)$, and retweet the meme that was present in their feed at that time (Fig.~\ref{fig:memory}).

In all the following analysis we shall consider the spreading of a meme from the node $i$. Now, the likelihood of a given meme being retweeted by the account is dependent on how quickly other memes enter their stream as the more content an account receives, the less likely they are to retweet any one meme. We now consider the different ways in which a meme may be accepted into the stream of account $i$. Firstly, account $i$ may innovate themselves (the new meme immediately appears in their stream), which occurs at a rate $\mu_i \beta_i$.  Secondly, account $i$ receives content from the individuals whom they follow on the same social media platform i.e.,~on the same layer $\alpha$; the rate at which such memes are received depends on the number of accounts followed, how active these individuals are and also how interesting account $i$ finds their content, being given by $\sum_{k}\lambda_{ki}\beta_k$. Account $i$ may also share content that they themselves have seen on another one of their social media platforms, this rate depends on the number of platforms the user is present on, how active they are on these platforms and also how likely they are to share from one medium to another, and is given by  $\sum_k c_{ki}\beta_k$. Finally, the individuals whom they follow on their own layer $\alpha$ may also share content that they seen on another of their accounts which our user may or may not find interesting, this occurs with rate $\sum_{k}\left[\lambda_{ki}\left(\sum_{l}c_{lk}\beta_l\right)\right]$. New memes thus enter the stream of user $i$ as a Poisson process with constant rate given by the sum of the rates described above:

\begin{equation}
r_i = \mu_i\beta_i + \sum_{k}^{}\lambda_{ki}\beta_k + \sum_{k}c_{ki}\beta_k + \sum_{k}\left[\lambda_{ki}\left(\sum_{l}c_{lk}\beta_l\right)\right].
\end{equation}
 Consequently, the time for which a single meme occupies the stream of account $i$ is an exponentially distributed random variable with probability density
\begin{equation}
P_{\text{occ},i}(\ell) = r_i\exp(-r_i\ell). \label{eq2}
\end{equation}

Now, as in \cite{gleeson2016effects}, we consider the size of retweet trees observed at a time $\O$, which arise from the successful insertion of a meme at time $\tau$ into the stream of account $i$. The tree would be created by the retweeting by account $i$ at some time point between $\tau$ and $\O$ as a result of looking back in their stream to a time $r$ when the meme was present, i.e., in the interval bounded by $\tau$ and $\min(\tau+\ell, \O)$. We now consider the small time interval $\dd r$ and the size of trees that occur as a result of looking back into this interval from a time $t$. The probability of looking back in this way will be dependent on the individual traits of the user on account $i$, specifically their activity rate $\beta_i$, their innovation rate $\mu_i$ and also their memory distribution $\Phi_i$. The probability that a tree is seeded at time $\dd t$ by retweeting the meme that was present in the time interval $\dd r$ is given (similar to Eq.~(5) of \cite{gleeson2016effects}) by

\begin{equation}
P_{\text{seed}, i} = (1-\mu_i) \beta_i \Phi_i(t-r)\,\dd t\,\dd r.
\end{equation}

We are now interested in determining how the relationship between two accounts affects the popularity of a meme which enters one of their streams. We define $G_{ij}(\tau,\O;x)$ to be the probability generating function (pgf) \cite{wilf2005generatingfunctionology} for the excess popularity of a meme, which entered account $i$'s stream at time $\tau$, as a result of their link to account $j$, observed at time $\O$:
\begin{equation}
G_{ij}(\tau,\O;x) = \sum_{n = 0}^{\infty}q_{ij, n}(\tau, \O) x^n,
\end{equation}
where $q_{ij,n}(\tau,\O)$ is the probability that a meme which entered user $i$'s feed at time $\tau$ has received $n$ retweets by time $\O$ as a result of the edge between nodes $i$ and $j$. 
 We note that this suggests a tree-based approximation of the network which is assumed in the model due to the independence of the branching processes
\cite{melnik2011unreasonable}.

Considering the distribution of excess tree sizes from a meme that has been seeded at time $t$ we note that once a tree has been seeded it is immediately seen by the followers of account $i$ on the same layer. An account $j$ on this layer may find the content interesting with probability $\lambda_{ij}$ and thus may retweet it in the future. If the account $j$ belongs to the same user as account $i$, but on a different layer, the meme may be shared to this site with  probability $c_{ij}$ and is then instantaneously transmitted to all of their followers, each of whom may find it interesting with probability $\lambda_{jl}$. Thus the pgf for tree size at $\O$ is\footnote{As the total excess tree size on any layer would be the sum of the tree sizes for those who found the content interesting, we multiply all their excess tree pgfs (since the pgf for the sum of random variables is given by the product of the pgfs corresponding to each variable).}
\begin{equation}
R_{ij}(t,\O;x) = x\left(1-\lambda_{ij} + \lambda_{ij}\prod_{k}^{}G_{jk}(t,\O;x)\right)\left\{1 - c_{ij} + c_{ij}\left[\prod_{l}\left(1 - \lambda_{jl} + \lambda_{jl}\prod_{m}G_{lm}(t,\O;x)\right)\right]\right\}. \label{e5}
\end{equation}

To determine the total tree size that may result from the account $i$ copying in the $\dd r$ interval and sharing with the account $j$ we must consider all times $t$ at which we may copy from and thus the total tree size from sharing from account $i$ to account $j$ is distributed by
\begin{equation}
J_{ij}(r;x) = \prod_{t = r}^{\O}\left[1 - P_{\text{seed}, i} + P_{\text{seed}, i}R_{ij}(t,\O;x)\right],
\end{equation}
which (as in Eq.~(9) of \cite{gleeson2016effects}) may be approximated by
\begin{equation}
J_{ij}(r;x) \rightarrow \exp\left\{-(1-\mu_i)\beta_i \,\dd r \int_{r}^{\O} \Phi_i(t-r)\left[1 - R_{ij}(t,\O;x)\right]\,\dd t \right\} \quad \text{ as } \dd t \to 0.
\end{equation}
As this is the size of a tree resulting from looking back to the time $\dd r$, to determine the total tree size culminating from the meme being present in the stream of account $i$ we must consider all times $r$ from $\tau$ to $\min(\tau + \ell, \O)$, so that the total size pgf is given by
\begin{align}
P_{\text{size},ij}(\ell) &= \prod_{r = \tau}^{\min(\tau+\ell,\O)} J_{ij}(r;x) \nonumber\\
&\rightarrow \exp\left\{-(1-\mu_i)\beta_i \int_{\tau}^{\min(\tau+\ell,\O)}\,\dd r \ \int_{r}^{\O} \Phi_i(t-r)\left[1 - R_{ij}(t,\O;x)\right]\,\dd t \right\} \quad \text{ as } \dd r \to 0.
\end{align}

Considering these probabilities for all the possible occupation times $\ell$ of the meme in the stream, as given by Eq.~(\ref{eq2}), we integrate to obtain
\begin{align}
G_{ij}(\tau, \O; x) &= \int_{0}^{\infty}P_{\text{occ}, i}(\ell)P_{\text{size}, ij}(\ell) \ \dd \ell \nonumber \\
&= \int_{0}^{\infty}r_i\exp{\left(-r_i \ell\right)}\exp\left\{\vphantom{\int_{\tau}^{\min(\tau+\ell,\O)}}-(1-\mu_i)\beta_i\right. \nonumber\\
&\left. \qquad \quad \times \int_{\tau}^{\min(\tau+\ell,\O)}\dd r \ \int_{r}^{\O} \Phi_i(t-r)\left[1 - R_{ij}(t,\O;x)\right]\ \dd t\right\} \ \dd \ell.
\end{align}
Introducing the change of variables $a = \O - \t, \ \widetilde{r} = r - \t, \ \widetilde{t} = \O -t$, we obtain
\begin{align}
G_{ij}(\O - a, \O; x) &= \int_{0}^{\infty}r_i\exp{\left(-r_i \ell\right)}\exp\left\{-\vphantom{\int_{\tau}^{\min(\tau+\ell,\O)}}(1-\mu_i)\beta_i \right. \nonumber\\
& \left. \qquad \quad \times \int_{0}^{\min(\ell,a)}\dd \widetilde{r} \ \int_{0}^{a-\widetilde{r}} \Phi_i(a-\widetilde{r}-\widetilde{t})\left[1 - R_{ij}(\O-\widetilde{t},\O;x)\right]\ \dd \widetilde{t}\right\} \ \dd \ell.
\end{align}
We note, as in \cite{gleeson2016effects}, that the only appearance of $\O$ in the above is in the first two arguments of $G$, and as such the popularity of memes depends only on their age rather than the global time. Therefore we may define $G_{ij}$ in terms of the age of the meme only, which allows us to write the closed equation
\begin{equation}
G_{ij}(a; x) = \int_{0}^{\infty}r_i\exp{\left(-r_i \ell\right)}\exp\left\{-(1-\mu_i)\beta_i \int_{0}^{\min(\ell,a)}\dd \widetilde{r} \ \int_{0}^{a-\widetilde{r}} \Phi_i(a-\widetilde{r}-\widetilde{t})\left[1 - R_{ij}(\widetilde{t};x)\right]\ \dd \widetilde{t}\right\} \ \dd \ell.
\label{Gpgf}
\end{equation}
This equation determines the pgf of the distribution of excess popularities at age $a$ as a result of the connection between user $i$ and $j$; we note that the pgfs describing both the entire excess distribution for a meme of age $a$ that entered user $i$'s feed $G_i(a;x)$ and also the distribution for popularities of a meme which was created by account $i$ via innovation $H_i(a;x)$, are easily calculated by following the corresponding derivations in \cite{gleeson2016effects}. For the analysis described in the remainder of this paper however Eq.~(\ref{Gpgf}) will be sufficient.

\section{Criticality of the Branching Process}\label{criticality}
The pgf described in Eq.~(\ref{Gpgf}) essentially describes an edge-specific quantity, namely the probability distribution of popularities based upon the link between account $i$ and account $j$, and as such this model may can be thought of as an age-dependent multi-type branching process \cite{athreya2004branching}. Recall that  in a classic single-type branching process the criticality of the system is determined by the `branching number' $\xi$, which is the mean number of `children' from each `parent': the branching process is exactly critical if the value of $\xi$ is 1, while it is subcritical if $\xi$ is less than 1 and supercritical in the case where $\xi$ exceeds 1.

In the theory of  multi-type branching processes \cite{athreya2004branching, harris2002theory} there is a similar measure of the system's criticality, determined by the the single reproductive number matrix, $\mathbf{M}$, with elements $m_{ij}$, which is the expected number of children of type $j$ that a parent of type $i$ produces. The long term behavior of such a process is then determined by the largest eigenvalue $\rho$ of the matrix  $\mathbf{M}$ such that $\rho < 1$ describes a subcritical process and $\rho > 1$ implies the process is supercritical. Again, the branching process is exactly critical when $\rho$ is 1.

As in Sec.IV.A of \cite{gleeson2016effects} we classify a meme that was accepted into account $i$'s stream at time $\tau$ as a parent of type $i$, and the retweets of it that are accepted into the stream of account $j$ at some time $t > \tau$ as the children of type $j$. The pgf for the number of children is derived in a similar manner to Eq.~(\ref{Gpgf}) but we replace $R_{ij}$ with $\left(1-\lambda_{ij} + \lambda_{ij}x\right)\left(1 - c_{ij} + c_{ij}x\right)$, as now we are only interested in the meme being accepted into the feed of account $j$ and not what occurs thereafter; 
the resulting pgf is given by
\begin{align}
K_{ij}(a;x) &= \int_{0}^{\infty}r_i\exp{\left(-r_i \ell\right)}\exp\left\{-(1-\mu_i)\beta_i \int_{0}^{\min(\ell,a)}\dd \widetilde{r}\int_{0}^{a-\widetilde{r}} \Phi_i(a-\widetilde{r}-\widetilde{t})\right. \nonumber \\
&\left. \hspace{10em} \times\left[1 - (1-\lambda_{ij}+\lambda_{ij}x)(1-c_{ij}+c_{ij}x)\right]\ \dd \widetilde{t}\vphantom{\int_{0}^{\min(\ell,a)}} \ \right\} \ \dd \ell,
\end{align}
The expected number of children is then found by differentiating $K_{i j}$ with respect to $x$ the above and evaluating at $x = 1$:
\begin{align}
m_{ij} &= \left.\pd{K_{ij}}{x}\right|_{x=1} \nonumber \\
&= \int_{0}^{\infty}r_i\exp{\left(-r_i \ell\right)}\exp\left\{-(1-\mu_i)\beta_i \int_{0}^{\min(\ell,a)}\dd \widetilde{r}\int_{0}^{a-\widetilde{r}} \Phi_i(a-\widetilde{r}-\widetilde{t})(c_{ij}+\lambda_{ij})\ \dd \widetilde{t}\right\} \ \dd \ell,
\end{align}
now taking the large-$a$ limit we obtain
\begin{equation}
m_{ij} \sim \frac{(c_{ij} + \lambda_{ij})(1-\mu_i)\beta_i}{\mu_i\beta_i + \sum_{k}^{}\lambda_{ki}\beta_k + \sum_{k}c_{ki}\beta_k + \sum_{k}\left[\lambda_{ki}\left(\sum_{l}c_{lk}\beta_l\right)\right]} \quad\text{ as } a \to \infty.
\label{xi}
\end{equation}
\subsection{Analysis of the Mean Matrix}\label{mean_analysis}
To determine the criticality of this process we must evaluate the maximum eigenvalue of the matrix $\mathbf{M}$ with elements given by Eq.~(\ref{xi}). First we shall consider the case of a monoplex network with no innovation and determine the criticality of such a system, then we'll include the (small) innovation probability and using a perturbative analysis we shall analyze the effect this has on the criticality of our process. Finally, we'll discuss the multiplex case with both innovation and crossover probabilities, and we'll show that the behavior of this system is purely dependent on the layer of the mean matrix which has the largest eigenvalue.

\begin{itemize}[leftmargin=*] 
	\item \textit{Case 1: Monoplex with $\mu_i = 0$}\\
	In this case we are essentially using one of the layers represented by adjacency matrix $A_{\alpha}$. As such there is no crossover probability, i.e., $c_{ij} = 0, \ \forall\ i,j$ and Eq.~(\ref{xi}) reduces to
	\begin{equation}
	\overline{m}_{ij} = \frac{\lambda_{ij}\beta_i}{ \sum_{k}^{}\lambda_{ki}\beta_k},
	\label{xi_mono}
	\end{equation}
In 	Appendix~\ref{AppA} we prove that that the largest eigenvalue $\overline{\rho}$ of such a matrix is equal to 1, this implies that in the case of a monoplex network where there is no innovation, the branching process is critical. We note that this is in agreement with the result found in ref.~\cite{gleeson2016effects} which showed that the degree-class version of the model described in Sec.~\ref{model} also approached criticality in the limit as $\mu \to 0$. We also note that as the matrix $\overline{\mathbf{M}}$ is non-negative the Perron-Frobenius theorem \cite{lancaster1985theory} states that eigenvalue equal to the spectral radius of such a matrix has positive left and right eigenvectors $u$ and $v$. This result will be important for our arguments in the following cases.
	
	\item \textit{Case 2: Monoplex with $\mu_i \ge 0$}\\
	Taking into consideration now the possibility of innovation on a single layer, i.e., $\mu_i \ge 0$ but $c_{ij} = 0, \ \forall\ i,j$, we note that Eq.~(\ref{xi}) is then given by
	\begin{equation}
	\widetilde{m}_{ij} = \frac{(1-\mu_i)\lambda_{ij}\beta_i}{\mu_i\beta_i + \sum_{k}^{}\lambda_{ki}\beta_k}.
	\label{xi_mono_inov}
	\end{equation}
 If we now consider the innovation probabilities to be small\footnote{We note that this assumption is motivated by empirical data in \cite{gleeson2016effects}, where the innovation probability for hashtags was estimated as $\mu = 0.055$.}, i.e., $\mu_i = \epsilon \widetilde{\mu}_i$ with $\widetilde{\mu}_i$ being $\mathcal{O}(1)$ and $\epsilon \ll 1$, Eq.~(\ref{xi_mono_inov}) may be expressed to $\mathcal{O}(\epsilon)$ as
	\begin{equation}
	\widetilde{m}_{ij} =  \frac{\lambda_{ij}\beta_i}{ \sum_{k}^{}\lambda_{ki}\beta_k} - \epsilon \, \frac{ \widetilde{\mu}_i\lambda_{ij}\beta_i\left(\beta_i + \sum_{k}\lambda_{ki}\beta_k \right)}{\left( \sum_{k}\lambda_{ki}\beta_k\right)^2}.
	\label{xi_difference}
	\end{equation}
So we write the matrix $\widetilde{\mathbf{M}} = \overline{\mathbf{M}} + \epsilon \Delta \overline{\mathbf{M}}$, and let $\overline{\rho} + \Delta \overline{\rho}$ be the largest eigenvalue of $\widetilde{\mathbf{M}} $. The change in eigenvalue $\Delta \overline{\rho}$ can be estimated when $\epsilon \ll 1$ via a first order approximation as \cite{milanese2010approximating}
	\begin{equation}
	\Delta \overline{\rho} = \left.\epsilon \, \overline{\rho}\,'\,\right|_{\overline{\mathbf{M}}} = \frac{u^T\epsilon \Delta \overline{\mathbf{M}}v}{u^Tv},
	\label{delta_rho}
	\end{equation}
	where $\left.\overline{\rho}\,'\,\right|_{\overline{\mathbf{M}}}$ represents the derivative of $\overline{\rho} + \Delta \overline{\rho}$ evaluated at $\epsilon = 0$.
	Noting that $\Delta \overline{\mathbf{M}}$ is the matrix with all-negative elements $\frac{-\widetilde{\mu}_i\lambda_{ij}\beta_i\left(\beta_i + \sum_{k}\lambda_{ki}\beta_k\right)}{{\left( \sum_{k}\lambda_{ki}\beta_k\right)^2}}$ in Eq.~(\ref{delta_rho}), and that $u$ and $v$ are positive eigenvectors, we see that $\Delta \overline{\rho}$ as given in Eq.~(\ref{delta_rho}) is necessarily negative. Thus the largest eigenvalue decreases from 1 when there is a non-zero possibility of innovation, which implies that the branching process becomes subcritical in the case where an individual may innovate.

The change in eigenvectors associated with the largest eigenvalue as a result of the perturbation may also be calculated to give
	\begin{equation}
	\Delta v \approx \epsilon \left(\Delta \overline{\mathbf{M}}\right) v, \quad \Delta u \approx \epsilon \left(\Delta \overline{\mathbf{M}}\right)^Tu.
	\end{equation}
	
	\item \textit{Case 3: Multiplex with $\mu_i, c_{ij} \ge 0$}\\
	Considering now the entire model over a multiplex with non-zero innovation and crossover probabilities, Eq.~(\ref{xi}) may be expressed as
	\begin{equation}
	m_{ij} = \frac{(1-\mu_i)\lambda_{ij}\beta_i}{\mu_i\beta_i + \sum_{k}^{}\lambda_{ki}\beta_k} + \frac{c_{ij}\left[\beta_i(1-\mu_i)\left(\mu_i\beta_i+\sum_{k}\lambda_{ki}\beta_k\right)\right] - \lambda_{ij}\left[\beta_i(1-\mu_i)\left(\sum_{k}\left\{c_{ki}\beta_k + \lambda_{ki}\sum_{l}c_{lk}\beta_l\right\}\right)\right]}
	{\left(\mu_i\beta_i + \sum_{k}^{}\lambda_{ki}\beta_k\right)\left(\mu_i\beta_i + \sum_{k}^{}\lambda_{ki}\beta_k + \sum_{k}\left\{c_{ki}\beta_k + \lambda_{ki}\sum_{l}c_{lk}\beta_l\right\}\right)}.
	\label{xi_multi}
	\end{equation}
	We immediately note that the first term in the above is the matrix described by Eq.~(\ref{xi_mono_inov}), which we have shown has maximum eigenvalue less than one. Now let's consider how the second term on the right-hand-side of Eq.~(\ref{xi_multi}) affects the structure of the matrix $\mathbf{M}$. Again we assume the crossover probabilities to be small such that $c_{ij} = \epsilon \, \widetilde{c}_{ij}$ with $\widetilde{c}_{ij}$ being $\mathcal{O}(1)$ and $\epsilon \ll 1$. The elements with non-zero $\lambda_{ij}$ are on the diagonal blocks of the matrix such that each block, $\widetilde{\mathbf{M}}_{\alpha}$, represents a layer $\alpha$ of the multiplex. Now as $\epsilon$ is increased from zero the largest eigenvalue for each of these blocks will also decrease  from 1, which can be seen by using a similar argument as in the case of a monoplex with innovation but where the matrix is perturbed (to first order) by
	\begin{equation}
	\frac{- \epsilon \, \lambda_{ij}\sum_{k}\left\{\widetilde{c}_{ki}\beta_k + \lambda_{ki}\sum_{l}\widetilde{c}_{lk}\beta_l\right\}}{\left( \sum_{k}\lambda_{ki}\beta_k\right)^2}.
	\label{pert_m_c}
	\end{equation}
	If this matrix only consisted of these diagonal blocks such that $\mathbf{M} = \oplus_{\alpha}\widetilde{\mathbf{M}}_\alpha$, the matrix would have the same set of eigenvalues as $\left\{\widetilde{\mathbf{M}}_\alpha\right\}$ and we could guarantee that the system is subcritical. However we also have the components in Eq.~(\ref{xi_multi}) which are grouped with the $c_{ij}$ terms and these are all off-diagonal entries which may modify the maximum eigenvalue of the matrix. Letting $\rho = \rho_\alpha + \epsilon\Delta\rho$ where $\rho_\alpha$ is the largest eigenvalue of $\oplus_{\alpha}\widetilde{\mathbf{M}}_\alpha$ and considering $\mathbf{M} = \oplus_{\alpha}\widetilde{\mathbf{M}}_\alpha + \epsilon \, \Delta \widetilde{\mathbf{M}}$, where $\Delta \widetilde{\mathbf{M}}$ is the matrix with elements (to first order) given by $\frac{\widetilde{c}_{ij}\beta_i}{\sum_{k}\lambda_{ki}\beta_k}$, we can determine the criticality of this system using $\rho_\alpha$. Again the change in the maximum eigenvalue (and its associated left and right eigenvectors) can be approximated to first order as
	\begin{equation}
	\Delta\rho = \frac{u^T \, \Delta \widetilde{\mathbf{M}} \, v}{u^Tv}, \quad \Delta v = \frac{\Delta \widetilde{\mathbf{M}}\,v}{\rho_\alpha}, \quad \Delta u = \frac{\left(\Delta \widetilde{\mathbf{M}}\right)^T\,u}{\rho_\alpha},\label{e22}
	\end{equation}
	 where $v$ ($u$) is the right (left) eigenvector associated with $\rho_\alpha$. Similarly to the approach of \cite{cozzo2013contact} we consider here the case of two layers $(\alpha = 1,2)$, but the argument may be easily generalized to an arbitrary number of layers. The two scenarios we must consider are $\rho_1 \gg \rho_2$ ($\rho_2 \gg \rho_1$ follows the same line of reasoning) and $\rho_1 \simeq \rho_2$.
\begin{itemize}
\item	When $\rho_1 \gg \rho_2$, layer one has the dominant eigenvalue with associated eigenvectors 
	\begin{equation}
	v = \left(\begin{array}{c}
	v_{(1)} \\ 
	0	\end{array}\right),\ \quad u = \left(\begin{array}{c}
	u_{(1)} \\
	0	\end{array}\right),
	\end{equation}
	and therefore the change in maximum eigenvalue given by Eq.~(\ref{e22}) is $\Delta\rho = 0$. This implies that the criticality of the branching process is determined by the layer of the multiplex which is itself most near to criticality.
In this sense the existence of a dominant layer guarantees that the system is subcritical.
	
\item	In the case where $\rho_1 = \rho_2$ the eigenvectors related to the maximum eigenvalue of the matrix are given by
	\begin{equation}
	v = \left(\begin{array}{c}
	v_{(1)} \\
	v_{(2)}	\end{array}\right),\ \quad u = \left(\begin{array}{c}
	u_{(1)} \\
	u_{(2)}	\end{array}\right),
	\end{equation}
	and therefore the change in eigenvalue is given by
	\begin{equation}
	\Delta \rho = \frac{u_{(2)}\Delta \widetilde{\mathbf{M}}_{12}v_{(1)}+u_{(1)}\Delta \widetilde{\mathbf{M}}_{21}v_{(2)}}{u_{(1)}^Tv_{(1)} + u_{(2)}^Tv_{(2)}},
	\label{delta_rho_sim}
	\end{equation}
	where $\Delta \widetilde{\mathbf{M}}_{ab}$ describes the block in matrix $\Delta \widetilde{\mathbf{M}}$ which represents coupling from layer $a$ to $b$. We cannot make any definitive statement about the eigenvalue in this case, and so we resort to numerical simulations. 

\end{itemize}
\end{itemize}

\subsection{Numerical Simulation}\label{num_sim}
We now consider numerical simulations to validate the results obtained in Sec.~\ref{mean_analysis}, in these simulations we examine a multiplex composed of two layers ($M = 2$), each with $N = 10^5$ nodes obtained via a directed configuration-model with out-degree distribution given by $p_k$. To keep analysis as simple as possible we consider the zero-memory case introduced in \cite{gleeson2014simple} such that if an account decides to retweet they always choose the meme that is currently in their feed [$\Phi_i(t_m) = \delta(t_m)$] as well as homogeneous parameters among users such that $\mu_i = \mu, \ c_{ij} = c, \ \beta_i = \beta, \ \text{and} \ \lambda_{ij} = \lambda, \ \forall \ i,j$. Time units are also chosen in the simulations such that on average each user becomes active once per model time unit, i.e., $\beta = 1$. Finally, we also assume that each user has an account on both platforms, such that the coupling matrix $C$ consists only of identity matrices in both the off-diagonal blocks. 

\begin{figure}[b!]
	\centering
	\includegraphics[width=\linewidth]{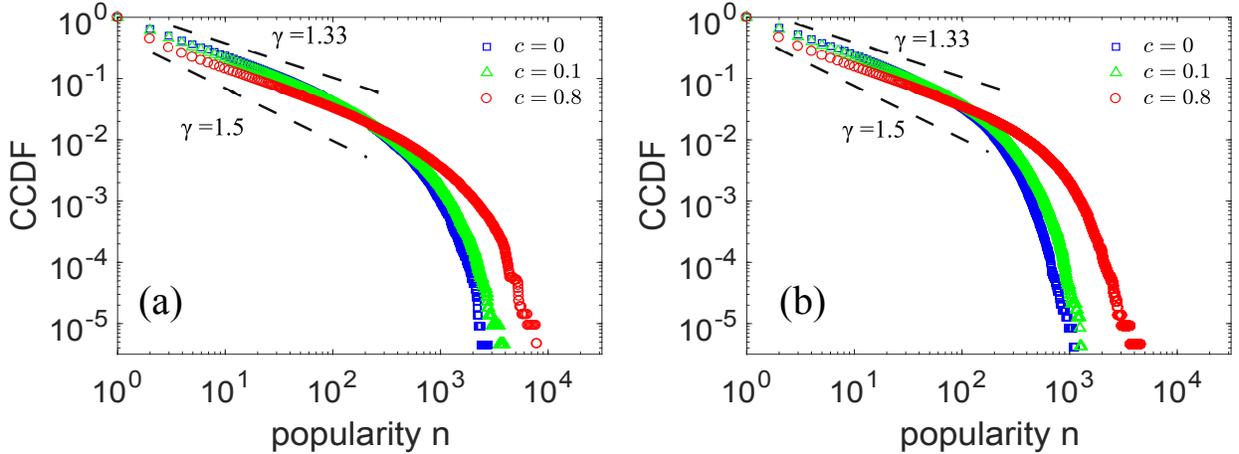}
	\caption{Complimentary Cumulative Distribution Functions for meme popularity at age 10 obtained via numerical simulations with $\mu = 0.05$ and multiple $c$ values, two different multiplexes are considered each with two layers of size $N = 10^5$. The multiplex used in (a) consists of layers with $p_k = \delta_{k,10}$ and $p_k = \delta_{k,2}$, while in (b) both layers have $p_k = \delta_{k,4}$. Dashed lines correspond to CCDFs for power law popularity distributions given by $n^{-\gamma}$.}
	\label{fig:ccdf}
\end{figure}

To validate the results of Sec.~\ref{mean_analysis} we consider two multiplexes, the first of which consists of layers where every account has exactly $k$ followers; $k = 10$ on the first layer and $k = 2$ on the second, which results in one of the layers having a dominant eigenvalue. The second multiplex consists of distinct networks on each layer both with $k = 4$, this ensures that the largest eigenvalues of each layer's mean matrix are very similar. To determine the effect of the crossover probability we perform simulations with $\mu = 0.05$ and $c  =  \{0, 0.1, 0.8\}$. The resulting CCDF of meme popularities at age 10 when the two layers have different out-degree distributions are shown in Fig.~\ref{fig:ccdf}(a) and when the layers have the same out-degree distribution in Fig.~\ref{fig:ccdf}(b), we immediately note that increased crossover probability results in larger cascades of retweet popularity which is justified by the fact that there are now more users who may retweet a meme when it appears on multiple platforms. 

\begin{table*}[t]
\renewcommand{\arraystretch}{1.5}
\centering
\begin{tabular}{|c|ccc|ccc|}
\cline{2-7}
\multicolumn{1}{c|}{} &  \multicolumn{3}{|c|}{$\rho_1 \gg \rho_2$} & \multicolumn{3}{|c|}{$\rho_1 \approx \rho_2$} \\
\hline
$p_k$ & & $\delta_{k,10}$ & $\delta_{k,2}$ & & $\delta_{k,4}$ & $\delta_{k,4}$ \\
\hhline{|=|===|===|}
$c$ & $\rho$ & $\rho_1$ & $\rho_2$ & $\rho$ & $\rho_1$ & $\rho_2$ \\
\hline
0&0.9889&0.9889&0.8508&0.9178&0.9178&0.9174 \\
$10^{-4}$&0.9887&0.9887&0.8506&0.9177&0.9177&0.9173 \\
$10^{-3}$&0.9878&0.9878&0.8459&0.9171&0.9166&0.9162 \\
$10^{-1}$&0.9000&0.8899&0.7361&0.8496&0.8122&0.8118 \\
0.8&0.5915&0.5236&0.3813&0.8941&0.4516&0.4514 \\
\hline
\end{tabular}
\caption{Leading eigenvalues, $\rho$, of the system's mean matrix, for the two multiplexes described in Sec.~\ref{num_sim}, where each layer has out-degree distribution given by $p_k$. Various crossover probabilities, $c$, are considered and in all cases $\mu = 0.05$. Also shown is the leading eigenvalue of each individual layer's mean matrix ($\rho_1, \rho_2$). We consider the case where there exists a dominant layer ($\rho_1 \gg \rho_2$) and also the contrary ($\rho_1 \approx \rho_2$).}
\label{rho_table}
\end{table*}

The numerical spectral analysis of this system for both multiplexes is also considered in Table \ref{rho_table}, which shows the largest eigenvalue of the whole system ($\rho$)\ in addition to each layer ($\rho_1, \rho_2$), for the crossover probabilities shown in Fig.~\ref{fig:ccdf}(a,b) as well as $c = \{10^{-4}, 10^{-3}\}$. The results from our perturbative analysis of the mean matrix in Sec.~\ref{mean_analysis} may now also be considered; Eq.~(\ref{pert_m_c}) suggests that with larger $c$ values, the maximum eigenvalue of each layer would decrease, which is which is consistent with the results in Table \ref{rho_table}. Secondly we note that in the case where there exists a dominant layer ($\rho_1 \gg \rho_2$), the spectral radius of the system's mean matrix is approximately equal to that of the dominant layer, particularly for smaller $c$ values, and in all cases the presence of intra-layer links moves the system closer to its critical point. The case where the mean matrices of both layers have similar leading eigenvalues ($\rho_1 \approx \rho_2$) may also be analyzed. In this case, for the smallest $c$ values $(0,10^{-4})$, the largest eigenvalue of the system is approximately equal to that of the dominant layer, however the perturbative effect noted by Eq.~(\ref{delta_rho_sim}) soon becomes apparent as the crossover probability increases, such that even for $c$ values as small as $10^{-3}$ the largest eigenvalue of the system is noticeably different from that of the dominant layer and in fact for crossover probabilities of much larger magnitude $(0.8)$ it appears that the system is much closer to its critical point than either of the layers individually.

Finally, we comment on the fact that in all examples described here the system is subcritical and we hypothesize that this system is in fact subcritical for all valid parameter values (while approaching criticality as the innovation probabilities tend toward zero). We hope that this result will encourage further research into the criticality of this system with the aim of validating this hypothesis.

\section{Discussion}\label{discussion}

In this paper we have extended the model for meme spreading that was introduced in \cite{gleeson2014competition, gleeson2016effects}. Instead of assuming that all nodes in each degree-class of the network behave the same, we here consider a specific network, defined by its adjacency matrix. In addition, we allow for each individual user $i$ to have his/her own parameter values, such as tweeting activity rate $\beta_i$. Moreover, we generalize beyond the monoplex (single-platform) case considered in \cite{gleeson2014competition, gleeson2016effects} to model the effects of users having accounts on multiple social media and potentially sharing information across platforms.

By developing an analytical approximation in terms of a multi-type branching process, we derive the equations (\ref{e5}) and (\ref{Gpgf}) for the probability generating functions of meme popularity. Although the complexity of these equations renders them intractable for large networks, we can nevertheless investigate the criticality of the branching process by spectral analysis of the mean matrix (Eq.~(\ref{xi})). We have two main results. First, we show that the criticality of the dynamical system for monoplex networks, as found in  \cite{gleeson2014competition, gleeson2016effects} for the vanishing-innovation limit, is robust, meaning that the multi-type branching process for a specific network (and with user-specific parameters) is also critical in the limit of zero innovation. Secondly, when we consider a multiplex network to model cross-platform transmission of information we show that the criticality of the system can be reduced to the spectral analysis of the mean matrix (Eq.~(\ref{xi})). In the case where the innovation probability is small, we show that if one layer of the multiplex has a dominant eigenvalue then the criticality of the multi-type branching process is determined by the layer of the multiplex that is closest to criticality. In the case where two layers have similar eigenvalues the analysis is more difficult, but numerical experiments suggest that the system is subcritical for all parameter values. We have also shown detailed numerical simulations of this model which results in heavy-tailed popularity distributions for a number of parameter values, and suggests that the potential for crossover between platforms results in larger cascades of meme popularity than on single-layer networks. Direct calculations of the spectral radius of the mean matrix in these simulations were also performed and further validated our results regarding the dependence of the system's criticality upon the existence (or not) of a dominant layer.

In conclusion, we believe that the generalization of the results of \cite{gleeson2014competition, gleeson2016effects} beyond random networks to demonstrate the possibility of near-critical dynamics in specific and multiplex networks is potentially important to the understanding of information spreading in real-world scenarios. Our main approximation is the assumption of branching-process dynamics within the model, but such approximations have been found to be quite accurate in similar dynamical systems on networks \cite{melnik2011unreasonable,Gleeson2017NatComm}. We hope that the proof of possible near-critical dynamics in this simple model of human behavior will inspire further theoretical and empirical work on information spreading in multiplex networks.

\section*{Acknowledgements}
This work was supported by Science Foundation Ireland grant numbers 16/IA/4470, 16/RC/3918 (J.D.O'B and J.P.G), and 15/IA/3074 (I.K.D). 

\bibliographystyle{unsrt}
\bibliography{multiplex}

\appendix

\section{Proof of Criticality of $\overline{\mathbf{M}}$}\label{AppA}
\textbf{Theorem} The largest eigenvalue of the matrix $\overline{\mathbf{M}}=[\overline{m}_{ij}]_{1\leq i,j \leq N}$, with elements
\[
\overline{m}_{ij}=
\frac{\beta_i\lambda_{ij}}{\sum_{k=1}^N\lambda_{ki}\beta_k}
,\quad i,j=1,2,...,N,
\]
is 1.
\\\\
\textbf{Proof.} If $z\in\mathbb{C}$ is eigenvalue of $\overline{\mathbf{M}}$, and $u=\left[\begin{array}{cccc}
u_1&
u_2&
\dots&
u_N

\end{array}\right]^T$ is the corresponding eigenvector we have:
\begin{equation}
(zI-\overline{\mathbf{M}})u=0_{N,1}. \label{eev}
\end{equation}
Defining the quantities $c_i$ by
\begin{equation}
c_i=\sum_{k=1}^N\lambda_{ki}\beta_k,\quad\forall i=1,2,...,N,
\end{equation}
we can write Eq.~(\ref{eev}) as
\[
\left[\begin{array}{cccc}

\frac{
\beta_1\lambda_{11}
}
{
c_1
}
-z&
\frac{
\beta_1\lambda_{12}
}
{
c_1
}&
\dots&
\frac{
\beta_1\lambda_{1N}
}
{
c_1
}\\

\frac{
\beta_2\lambda_{21}
}
{
c_2
}&
\frac{
\beta_2\lambda_{22}
}
{
c_2
}-z&
\dots&
\frac{
\beta_2\lambda_{N1}
}
{
c_2
}\\

\vdots&
\vdots&
\ddots&
\vdots\\

\frac{
\beta_N\lambda_{N1}
}
{
c_m
}&
\frac{
\beta_N\lambda_{N2}
}
{
c_n
}&
\dots&
\frac{
\beta_N\lambda_{NN}
}
{
c_N
}-z

\end{array}\right]
\left[\begin{array}{cccc}
u_1\\
u_2\\
\vdots\\
u_N

\end{array}\right]
=
\left[\begin{array}{cccc}
0\\
0\\
\vdots\\
0

\end{array}\right],
\]
or, equivalently,
\[
\left[\begin{array}{cccc}

\beta_1\lambda_{11}-zc_1
&
\beta_1\lambda_{12}
&
\dots&
\beta_1\lambda_{1N}
\\
\beta_2\lambda_{21}
&
\beta_2\lambda_{22}-zc_2
&
\dots&
\beta_2\lambda_{2N}
\\

\vdots&
\vdots&
\ddots&
\vdots\\

\beta_N\lambda_{N1}
&
\beta_N\lambda_{N2}
&
\dots&
\beta_N\lambda_{NN}-zc_N\frac

\end{array}\right]
\left[\begin{array}{cccc}
\hat{u}_1\\
\hat{u}_2\\
\vdots\\
\hat{u}_N

\end{array}\right]
=
\left[\begin{array}{cccc}
0\\
0\\
\vdots\\
0

\end{array}\right],
\]
where
\[
\hat{u}=\left[
\begin{array}{c}
\hat{u}_1\\
\hat{u}_2\\
\vdots\\
\hat{u}_N
\end{array}\right]
=
diag\Big\{\frac{1}{c_i}\Big\}_{1\leq i\leq N}
\left[\begin{array}{c}
u_1\\
u_2\\
\vdots\\
u_N
\end{array}\right].
\]
Defining $B$ as the matrix
\[
B(z)=\left[\begin{array}{cccc}

\beta_1\lambda_{11}-zc_1
&
\beta_1\lambda_{12}
&
\dots&
\beta_1\lambda_{1N}
\\
\beta_2\lambda_{21}
&
\beta_2\lambda_{22}-zc_2
&
\dots&
\beta_2\lambda_{2N}
\\

\vdots&
\vdots&
\ddots&
\vdots\\

\beta_N\lambda_{N1}
&
\beta_N\lambda_{N2}
&
\dots&
\beta_N\lambda_{NN}-zc_N\frac

\end{array}\right],
\]
we now suppose that  an eigenvalue with $|z|>1$ exists, and we will show that we arrive at a contradiction. Considering column $j$ of the matrix $B$, we observe that since $|z|$ is assumed to be greater than 1,
\begin{align*}
&\sum^N_{i\neq j}\beta_i\lambda_{ij} < |z| \sum^N_{i\neq j}\beta_i\lambda_{ij} <  |z| \sum^N_{i\neq j}\beta_i\lambda_{ij} + (|z|-1) \beta_j\lambda_{jj},\\
& \Longleftrightarrow \sum^N_{i\neq j}\beta_i\lambda_{ij} <  |z| \sum^N_{i= 1}\beta_i\lambda_{ij} - \beta_j\lambda_{jj}\\
& \Longleftrightarrow \sum^N_{i\neq j}\beta_i\lambda_{ij} <  |z|c_j - \beta_j\lambda_{jj}.
\end{align*}
Since all the $\beta_j$, $c_j$, $\lambda_{ij}$ are non-negative we have
\[
\sum^N_{i\neq j}\beta_i\lambda_{ij} <  |zc_j| - |\beta_j\lambda_{jj}| \le | zc_j -\beta_j\lambda_{jj} |.
\]
This implies that the matrix $\overline{\mathbf{M}}(z)$ is strictly diagonally dominant, i.e. $\overline{\mathbf{M}}(z)$ is non-singular (this can be proved by using the Gershgorin circle theorem, see \cite{golub2012matrix}). However, if  $\overline{\mathbf{M}}(z)$ is  non-singular then
$\hat{u}$, and consequently $u$, will be equal to $0_{N,1}$, contradicting the fact that $u$ is an eigenvector of $z$. Thus, based on assuming $|z|>1$ we have derived a contradiction and so we conclude that there do not exist eigenvalues with $|z|>1$.
\\\\
For $z=1$ the matrix $\overline{\mathbf{M}}(z)$ takes the form
\[
\overline{\mathbf{M}}(1)=\left[\begin{array}{cccc}

\beta_1\lambda_{11}-c_1
&
\beta_1\lambda_{12}
&
\dots&
\beta_1\lambda_{1N}
\\
\beta_2\lambda_{21}
&
\beta_2\lambda_{22}-c_2
&
\dots&
\beta_2\lambda_{2N}
\\

\vdots&
\vdots&
\ddots&
\vdots\\

\beta_N\lambda_{N1}
&
\beta_N\lambda_{N2}
&
\dots&
\beta_N\lambda_{NN}-c_N\frac

\end{array}\right].
\]
It is easy to observe that each column of $\overline{\mathbf{M}}(1)$ sums to zero, i.e. $\overline{\mathbf{M}}(1)$ is singular, and $z=1$ is an eigenvalue of $\overline{\mathbf{M}}$. This completes the proof.

\end{document}